\documentclass[twocolumn,aps]{revtex4}

\usepackage{graphicx,epic,eepic,epsfig,amsmath,latexsym,amssymb,verbatim}
\usepackage{theorem}

\begin{document}
\input epsf

\title{On the entanglement concentration of three-partite states}

\author{Berry Groisman$^a$, Noah Linden$^b$, and Sandu Popescu$^{a,c}$}
\affiliation{$^a$HH Wills Physics Laboratory, University of
Bristol, Tyndall Avenue, Bristol, BS8 1TL, UK
\\$^b$Department of Mathematics, University of Bristol, University
Walk, Bristol BS8 1TH, UK\\
$^c$Hewlett-Packard Laboratories, Stoke Gifford, Bristol BS12 6QZ,
UK }



\begin{abstract}
We investigate the concentration of multi-party entanglement by
focusing on simple family of three-partite pure states,
superpositions of Greenberger-Horne-Zeilinger states and singlets.
Despite the simplicity of the states, we show that they cannot be
reversibly concentrated by the standard entanglement concentration
procedure, to which they seem ideally suited. Our results cast
doubt on the idea that for each N there might be a finite set of
N-party states into which any pure state can be reversibly
transformed.  We further relate our results to the concept of
locking of entanglement of formation.

\end{abstract}


\maketitle

\section{Introduction}\label{intr}

Entanglement of bi-partite states is very well studied and for
pure states, the situation is particularly simple \cite{entconc}:
all bi-partite pure states are equivalent to singlets, as far as
their entanglement is concerned, in sense that $n$ copies of the
state are reversibly converted into singlets using Local
Operations and Classical Communication (LOCC) in the (asymptotic)
limit $n \rightarrow \infty$.

Multi-partite entanglement is much less well understood. It is
known that a general multi-partite non-maximally entangled state
cannot be reversibly concentrated into a collection of singlets
between pairs of parties \cite{LPSW,multiparty}. It was
conjectured, however, that in the asymptotic limit every pure
multi-partite entangled state can be reversibly converted into
combination of states from a certain Minimal Reversible
Entanglement Generating Set (MREGS), which plays the role of a
singlet state in a bi-partite case \cite{multiparty}.

In this paper we address the question of whether the MREGS
conjecture really works. We consider perhaps the simplest
non-trivial case - a special kind of non-maximally entangled
three-partite state that was first considered by D. Rohrlich
\cite{dr}:
\begin{eqnarray}\label{rstate}
|R_p\rangle=\sqrt{1-p}&|0\rangle_A\frac{|0\rangle_B|0\rangle_C+|1\rangle_B|1\rangle_C}{\sqrt{2}}~\nonumber\\
 +\sqrt{p}&|1\rangle_A\frac{|0\rangle_B|0\rangle_C-|1\rangle_B|1\rangle_C}{\sqrt{2}}.
 \end{eqnarray}
The state was analysed in more detail in \cite{LPSW}; in
particular, its asymptotic convertibility was discussed leading to
the open question addressed in this paper. It is natural to
conjecture that $n$ copies of $|R_p\rangle$ can be reversibly
concentrated into collection of $nH(p)$ GHZ-states \cite{GHZ} and
$n[1-H(p)]$ singlets held between $B$ and $C$ in the asymptotic
limit $n \rightarrow \infty$, where $H(p)=-p \log_2p-(1-p)
\log_2(1-p)$ is the (Shannon) entropy of the probability
distribution $\{p, 1-p\}$. In \cite{LPSW} evidence for this
conjecture was given based on the conservation of various
quantities in reversible procedures. It will also be seen below
that this proportion of GHZ's and singlets would result from the
standard method of entanglement concentration.

\cite{svw} contains interesting results for a number of related
questions, for example the optimal rate of extraction of
GHZ-states for $|R_p\rangle$ when the number of singlets extracted
is not important.

\section{The standard concentration method applied to multi-party states.}

The (standard) quantum entanglement concentration scheme
\cite{entconc} is inspired by the idea of classical Shannon
compression:  for example, the concentration of a large number $n$
of non-maximally entangled bi-partite states

\begin{equation}\label{first}
|\phi_p\rangle^{\otimes n} =[\sqrt
{1-p}|0\rangle_A|0\rangle_B+\sqrt
{p}|1\rangle_A|1\rangle_B]^{\otimes n}
\end{equation}

\noindent to a smaller number of maximally entangled states

\begin{equation}\label{EPR}
|\phi_{\frac{1}{2}}\rangle^{\otimes k}
=[\frac{1}{\sqrt{2}}(|0\rangle_A|0\rangle_B+|1\rangle_A|1\rangle_B)]^{\otimes
k}
\end{equation}

\noindent is based on the observation, that the total initial
state of $2n$ qubits, when expanded in local bases, contains {\it
typical} terms of the type

\begin{equation}\label{typical}
|0^{\otimes (n-k)}1^{\otimes k}\rangle_A
 |0^{\otimes (n-k)}1^{\otimes k}\rangle_B,
 \end{equation}

\noindent where $k$ has a value in the interval $[np \pm
O(\sqrt{n})]$. Here $0^{\otimes (n-k)}1^{\otimes k}$ denotes some
particular configuration of $k$ one's and $n-k$ zero's. If the
measurement of a {\it total number} of $1$'s on one of the sides
is performed than the result will most probably yield $k$ $1$'s,
where $k\in [np\pm O(\sqrt{n})]$. In this case the initial
non-maximally entangled state of $2n$ qubits will be projected to
a maximally entangled state, which contains ${n \choose k}$
orthogonal terms of the type (\ref{typical}). Thus, the state
which results if $k$ $1$'s is found is
\begin{equation}\label{global}
\frac{1}{\sqrt{{n \choose k}}}\sum_{i=1}^{{n \choose k}}
|P_i(0^{\otimes (n-k)}1^{\otimes k})\rangle_A
 |P_i(0^{\otimes (n-k)}1^{\otimes k})\rangle_B,
\end{equation}

\noindent where the sum runs over all possible permutations $P_i$
of $k$ one's and $n-k$ zero's. For large $n$ the value of ${n
\choose k}$ is approximated very well by $2^{nH(p)}$ \footnote{We
will use the mean value of $k$, i.e. $np$, keeping in mind the
deviations from it.}. In what follows we will often omit the
argument $p$ of the entropy $H(p)$ and will denote it just by $H$.

 In other words, by neglecting {\it atypical}
terms, which appear with very small probability, we have got a
maximally entangled state with Schmidt number $2^{nH}$.

This is, however, only part of the story, because the resulting
state is now an entangled state of all $2n$ particles which is not
partitioned into direct product of 2-particle maximally entangled
states. This is because the state (\ref{global}) ``lives" in a
$2^n\otimes2^n$-dimensional Hilbert space, which is spanned by
$2^n$ orthogonal states of $n$ qubits on each side. However,
(\ref{global}) contains only $2^{nH}$ orthogonal terms in Schmidt
decomposition and, in principle, can be ``compressed" to a
$2^{nH}\otimes2^{nH}$-dimensional Hilbert space. Thus, $n(1-H)$
qubits on each side are redundant.

 To make this explicit Alice and Bob each apply a collective local unitary
 transformation

\begin{eqnarray}\label{com}
 U|P_1(0^{\otimes (n-k)}1^{\otimes k})\rangle~~~~=|00...000\rangle |0\rangle^{\otimes
 n(1-H)}\nonumber\\
 U|P_2(0^{\otimes (n-k)}1^{\otimes k})\rangle~~~~=|00...001\rangle |0\rangle^{\otimes
 n(1-H)}\nonumber\\
 U|P_3(0^{\otimes (n-k)}1^{\otimes k})\rangle~~~~=|00...010\rangle |0\rangle^{\otimes
 n(1-H)}\\
.~~~~~~.~~~~~~.~~~~~~.~~~~~~.~~~~~~.~~~~~~.\nonumber\\
U|P_{2^{nH}}(0^{\otimes (n-k)}1^{\otimes
k})\rangle~=|11...111\rangle |0\rangle^{\otimes n(1-H)}\nonumber
 \end{eqnarray}
 \noindent on their particles, which re-arranges this state to a $2^{nH}\otimes2^{nH}$-dimensional subspace of the
 original Hilbert space and sets all redundant qubits to some standard state, e.g. the all $|0\rangle$-state.
 This local transformation is isomorphic to classical Shannon compression where $2^{nH}$ typical
 sequences which have a length $n$ are relabelled using $2^{nH}$ codewords which have length $nH$.
It is easy to check that the new state of $2nH$ qubits is nothing
but the direct product of $nH$ separate EPR-states, i.e.
(\ref{EPR}). Thus, as a result of (\ref{com}) the total state
(\ref{global}) of $2n$ particles is converted to

\begin{equation}
\biggl[\frac{1}{\sqrt{2}}(|0\rangle_A|0\rangle_B+|1\rangle_A|1\rangle_B)\biggr]^{\otimes
nH}\otimes \biggl[|0\rangle_A|0\rangle_B\biggr]^{\otimes n(1-H)}.
\end{equation}

\begin{figure}
\epsfxsize=3.3truein \centerline{\epsffile{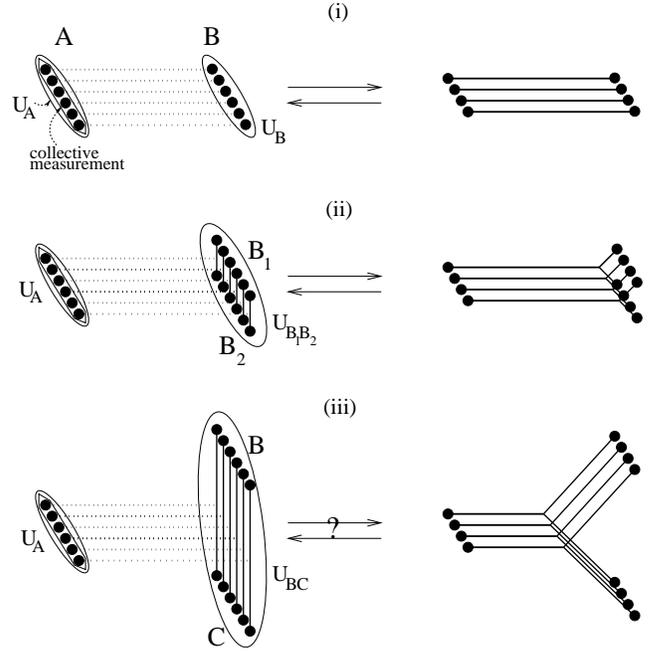}}
\caption[]{(i): Standard bi-partite entanglement concentration
scheme. Alice performs collective measurement on her side followed
by local unitary transformations $U_A$ and $U_B$. The process is
reversible in the asymptotic limit. (ii): The scheme works
similarly in the case when there are two particles (for each
state) on Bob's side. Bob performs the {\it local} unitary
transformation $U_{B_1B_2}$ on $2n$ qubits on his side. The
process is reversible in the asymptotic limit. (iii): $2n$ qubits
initially held by Bob are distributed now between Bob and Claire.
The question is whether Bob and Claire can perform $U_{BC}$ using
LOCC only. } \label{Fig1}
\end{figure}

Consider now the following {\it three-particle} bi-partite state
\begin{equation}\label{3local}
|\Phi_p\rangle^{\otimes n}
=(\sqrt{p}|0\rangle_A|{\theta}\rangle_{B_1B_2}+\sqrt{1-p}|1\rangle_A|{\tau}\rangle_{B_1B_2})^{\otimes
n}.
\end{equation}
Here $|{\theta}\rangle_{B_1B_2}$, $|{\tau}\rangle_{B_1B_2}$ are
normalised orthogonal but otherwise general states of two
particles $B_1$ and $B_2$ located on Bob's side. The entanglement
concentration procedure will work in this case just as before,
since Bob is able to apply all operations described above  working
locally with the states $|{\theta}\rangle_{B_1B_2}$,
$|{\tau}\rangle_{B_1B_2}$ of two particles exactly as he would
work with the states $|0\rangle_B$, $|1\rangle_B$ of one particle
(see Fig. \ref{Fig1}, (ii)). As a result, Alice and Bob will be
able reversibly to  concentrate $n$ copies of (\ref{3local}) into
$nH$ copies of the maximally entangled state
\begin{equation}
|\Phi_{\frac{1}{2}}\rangle
=\frac{1}{\sqrt{2}}(|0\rangle_A|{\theta}\rangle_{B_1B_2}+|1\rangle_A|{\tau}\rangle_{B_1B_2}),
\end{equation}
while the other $n(1-H)$ initially non-maximally entangled
``triples" are now in the state
$|0\rangle_A|{\theta}\rangle_{B_1B_2}$. Thus the total state is
\begin{eqnarray}
\biggl[\frac{1}{\sqrt{2}}(|0\rangle_A|{\theta}\rangle_{B_1B_2}+|1\rangle_A|{\tau}\rangle_{B_1B_2})\biggr]^{\otimes
nH}~~~~~~~~~\\
~~~~~~~~~~~~~~~~~~~\otimes\biggl[|0\rangle_A|{\theta}\rangle_{B_1B_2}\biggr]^{\otimes
n(1-H)}.\nonumber
\end{eqnarray}

Now suppose that initially Bob gives one of his particles to
Claire:

\begin{equation}\label{3nonlocal}
|\Psi_p\rangle^{\otimes n}
=(\sqrt{p}|0\rangle_A|{\theta}\rangle_{BC}+\sqrt{1-p}|1\rangle_A|{\tau}\rangle_{BC})^{\otimes
n}\end{equation}

Starting the procedure in the same way as we did in the case of
(\ref{3local}) we will soon get the following analog of
(\ref{global}):

 \begin{equation}\label{3comp1}
 \frac{1}{\sqrt{2^{nH}}}\sum_i|P_i(0^{\otimes (n-k)}1^{\otimes k})\rangle_A
 |P_i({\theta}^{\otimes (n-k)}{\tau}^{\otimes k})\rangle_{BC},
 \end{equation}

 \noindent where the sum runs over all possible permutations $P_i$.
 From the point of view of $A$ the procedure continues from here
 in the same way, namely Alice can apply local transformation
 (\ref{com}) on her local qubits. However, $B$ and $C$ must {\it jointly} apply
 a bi-partite analog $U_{BC}$ of (\ref{com}) on their qubits. This
 transformation may be written simply by replacing $|0\rangle$, $|1\rangle$
 with $|{\theta}\rangle$, $|{\tau}\rangle$ in (\ref{com}):
\begin{eqnarray}\label{comBC}
 |P_1({\theta}^{\otimes (n-k)}{\tau}^{\otimes k})\rangle~~~~\rightarrow|{\theta}{\theta}...{\theta}{\theta}{\theta}\rangle |{\theta}^{\otimes
 n(1-H)}\rangle\nonumber\\
 |P_2({\theta}^{\otimes (n-k)}{\tau}^{\otimes k})\rangle~~~~\rightarrow|{\theta}{\theta}...{\theta}{\theta}{\tau}\rangle |{\theta}^{\otimes
 n(1-H)}\rangle\nonumber\\
 |P_3({\theta}^{\otimes (n-k)}{\tau}^{\otimes k})\rangle~~~~\rightarrow|{\theta}{\theta}...{\theta}{\tau}{\theta}\rangle |{\theta}^{\otimes
 n(1-H)}\rangle\\
.~~~~~~.~~~~~~.~~~~~~.~~~~~~.~~~~~~.~~~~~~.\nonumber\\
|P_{2^{nH}}({\theta}^{\otimes (n-k)}{\tau}^{\otimes
k})\rangle\rightarrow|{\tau}{\tau}...{\tau}{\tau}{\tau}\rangle
|{\theta}^{\otimes n(1-H)}\rangle\nonumber
 \end{eqnarray}

As a side effect of this transformation, $2^{n(1-H)}$ redundant
pairs of $B$ and $C$ will be in the state $|{\theta}\rangle$.
Thus, if $U_{BC}$, (\ref{comBC}), can be performed then we should
get:

\begin{equation}
\biggl[\frac{1}{\sqrt{2}}(|0\rangle_A|{\theta}\rangle_{BC}+|1\rangle_A|{\tau}\rangle_{BC})\biggr]^{\otimes
nH}\otimes\biggl[|0\rangle_A|{\theta}\rangle_{BC}\biggr]^{\otimes
n(1-H)}.
\end{equation}

The main question is whether $U_{BC}$ can be achieved by LOCC.
Even if not, it might be achievable using an amount of
entanglement per copy which goes to zero as $n \rightarrow
\infty$. In this latter case we may use some singlets, but a
number which is negligible in the asymptotic limit, and thus the
transformation is still reversible. Thus, we ask what is the
minimal amount of entanglement between $B$ and $C$ needed to
implement $U_{BC}$ using this entanglement and LOCC.

{\bf Example I:} As a first example let us consider the case
when\\
\parbox{0.7in}{\begin{eqnarray*}\end{eqnarray*}}\hfill
\parbox{1.5in}{\begin{eqnarray*} |{\theta}\rangle_{BC}&=&|0\rangle_B|0\rangle_C~\\
|{\tau}\rangle_{BC}&=&|1\rangle_B|1\rangle_C
\end{eqnarray*}}\hfill
\parbox{0.7in}{\begin{eqnarray}\end{eqnarray}}\\
\noindent In this case (\ref{3nonlocal}) will correspond to
non-maximally entangled GHZ-states:
\begin{equation}
 |\Theta_p\rangle^{\otimes n}
=(\sqrt{1-p}|0\rangle_A|0\rangle_B|0\rangle_C+\sqrt{p}|1\rangle_A|1\rangle_B|1\rangle_C)^{\otimes
n} \end{equation}

If $U_{BC}$ can be performed then we should get $nH$ copies of
GHZ-state\\
\parbox{0.2in}{\begin{eqnarray*}\end{eqnarray*}}\hfill
\parbox{1.7in}{\begin{eqnarray*} \frac{1}{\sqrt{2}}(|0\rangle_A
|{\theta}\rangle_{BC}+|1\rangle_A|{\tau}\rangle_{BC}\nonumber)~~~~~~~~~~~
\\
=\frac{1}{\sqrt{2}}(|0\rangle_A|0\rangle_B|0\rangle_C+|1\rangle_A|1\rangle_B|1\rangle_C)
\end{eqnarray*}}\hfill
\parbox{0.7in}{\begin{eqnarray}\end{eqnarray}}\\
\noindent and $n(1-H)$ copies of
$|0\rangle_A|{\theta}\rangle_{BC}=|0\rangle_A|0\rangle_B|0\rangle_C$:

\begin{eqnarray}
\biggl[\frac{1}{\sqrt{2}}(|0\rangle_A|0\rangle_B|0\rangle_C+|1\rangle_A|1\rangle_B|1\rangle_C)\biggr]^{\otimes
nH}~~~~~~~~~\\
~~~~~~~~~~~~~~~~~~~\otimes\biggl[|0\rangle_A|0\rangle_B|0\rangle_C\biggr]^{\otimes
n(1-H)}.\nonumber
\end{eqnarray}
\noindent Clearly there is a possible bi-partite unitary $U_{BC}$
of the form $U_B \otimes U_C$ ($U_B$ and $U_C$ being of the form
(\ref{com})) which may be implemented locally by Bob and Claire.
Thus, $n$ non-maximally entangled GHZ-states can be reversibly
concentrated to $nH$ maximally entangled GHZ-states plus $n(1-H)$
direct products.

{\bf Example II:} We now arrive at the main point. We consider the
following choices of bi-partite states (introduced in
\cite{dr})\\
\parbox{0.4in}{\begin{eqnarray*}\end{eqnarray*}}\hfill
\parbox{2.in}{\begin{eqnarray*}
|{\theta}\rangle_{BC}=\frac{1}{\sqrt{2}}(|0\rangle_B|0\rangle_C+|1\rangle_B|1\rangle_C)~\nonumber\\
|{\tau}\rangle_{BC}=\frac{1}{\sqrt{2}}(|0\rangle_B|0\rangle_C-|1\rangle_B|1\rangle_C).
\end{eqnarray*}}\hfill
\parbox{0.5in}{\begin{eqnarray}\label{map}\end{eqnarray}}

\noindent This leads to what we call the Rohrlich state
(\ref{rstate}), which now can be rewritten in terms of
$|{\theta}\rangle$ and $|{\tau}\rangle$ as
\begin{eqnarray}\label{Tstate}
|R_p\rangle=\sqrt{1-p}|0\rangle_A|{\theta}\rangle_{BC}
 +\sqrt{p}|1\rangle_A|{\tau}\rangle_{BC}
 \end{eqnarray}
\noindent We note that the state $|R_{1 \over 2}\rangle$ is
locally equivalent to a GHZ state, and the state $|R_0\rangle$
comprises a singlet held between Bob and Claire.

If $U_{BC}$ could be implemented locally in this case  then $n$
copies of $|R_p\rangle$ would be reversibly converted (in the
asymptotic limit) into $nH$ GHZ states and $n(1-H)$ singlets
between Bob and Claire:

\begin{eqnarray}\label{final}
\biggl[\frac{1}{\sqrt{2}}(\frac{|0\rangle_A+|1\rangle_A}{\sqrt{2}~~}|0\rangle_B|0\rangle_C+\frac{|0\rangle_A-|1\rangle_A}{\sqrt{2}~~}|1\rangle_B|1\rangle_C)\biggr]^{\otimes
nH}\nonumber\\
\otimes
\biggl[|0\rangle_A\frac{1}{\sqrt{2}}(|0\rangle_B|0\rangle_C+|1\rangle_B|1\rangle_C)\biggr]^{\otimes
   n(1-H)};~~~~~~~
\end{eqnarray}
which is consistent with the conjecture in \cite{LPSW}.

In this paper we show that using the standard concentration
procedure such a transformation is impossible. Our conclusion
follows from examination of the amount of non-locality in
$U_{BC}$: we find that it is not negligible, but proportional to
$n$ as $n \rightarrow \infty$.

Here we use the following method in order to find the amount of
non-locality in $U_{BC}$ (see also \cite{Toffoli}). We act with
$U_{BC}$ on a {\it test state} $|\Psi_{in}^{test}\rangle_{BC}$:

\begin{equation}
U_{BC}|\Psi_{in}^{test}\rangle=|\Psi_{out}^{test}\rangle,
\end{equation}

\noindent where the test state is a superposition of basic input
states in (\ref{comBC}).
 We denote the
amount of non-locality between $B$ and $C$ possessed by
$|\Psi_{in}^{test}\rangle_{BC}$ and
$|\Psi_{out}^{test}\rangle_{BC}$ by $E_{in}^{test}$ and
$E_{out}^{test}$ respectively, where
$E=S(Tr_B|\Psi\rangle\langle\Psi|)=S(Tr_C|\Psi\rangle\langle\Psi|)$
is the von Neumann entropy of the reduced density matrix. If
$|\Psi_{in}^{test}\rangle_{BC}$, $|\Psi_{out}^{test}\rangle_{BC}$
possess different amount of entanglement, then $U_{BC}$ is
non-local. The amount of non-locality in $U_{BC}$ is not less than
the entanglement difference between the two states: $E_U \geq
|E_{in}^{test}-E_{out}^{test}|$ (acting on different test states
$U_{BC}$ may produce different amounts of entanglement).

\section{An example: $n$=4.}\label{4}
 It turns out that in the non-asymptotic case of $n=2$ this
 transformation can be implemented by LOCC. Indeed,
 \begin{eqnarray}
|{\theta}\rangle|{\tau}\rangle \rightarrow |{\theta}\rangle|{\theta}\rangle\nonumber\\
|{\tau}\rangle|{\theta}\rangle \rightarrow
|{\tau}\rangle|{\theta}\rangle\nonumber
 \end{eqnarray}
is nothing but a partial CNOT transformation on logical bits
encoded nonlocally in the states $|{\theta}\rangle$,
$|{\tau}\rangle$ followed by a NOT on the second logical qubit. It
can be easily checked explicitly that this nonlocal CNOT
transformation can be built from local CNOT gates.

However, for $n>2$ this transformation cannot be implemented by
LOCC \cite{Toffoli}. Let us illustrate this for $n=4$. Consider
the case of a single $|{\tau}\rangle$. Here two qubits are
redundant on each side and $U_{BC}$ maps the four possible terms
as follows
\begin{center}
\begin{eqnarray}\label{basic}
U_{BC}|{\theta}{\theta}{\theta}{\tau}\rangle=|{\theta}{\theta}{\theta}{\theta}\rangle~\nonumber\\
U_{BC}|{\theta}{\theta}{\tau}{\theta}\rangle=|{\theta}{\tau}{\theta}{\theta}\rangle~\\
U_{BC}|{\theta}{\tau}{\theta}{\theta}\rangle=|{\tau}{\theta}{\theta}{\theta}\rangle~\nonumber\\
U_{BC}|{\tau}{\theta}{\theta}{\theta}\rangle=|{\tau}{\tau}{\theta}{\theta}\rangle
,\nonumber
\end{eqnarray}\end{center}
i.e. two last pairs are in the $|{\theta}\rangle$-state while two
first pairs carry the information about the four possible inputs.

It is useful to consider the action of $U_{BC}$, defined in
(\ref{basic}), on a superposition. In particular, $U_{BC}$ will
transform the following test state
\begin{eqnarray}\label{in}
|\Psi_{in}^{test}\rangle_{BC}={1
\over2}\biggl(|{\theta}{\theta}{\theta}{\tau}\rangle_{BC}+|{\theta}{\theta}{\tau}{\theta}\rangle_{BC}
~~~~~~~\nonumber\\
+|{\theta}{\tau}{\theta}{\theta}\rangle_{BC}+|{\tau}{\theta}{\bf
}{\theta}{\theta}\rangle_{BC}\biggr)
\end{eqnarray}
to the state
\begin{eqnarray}\label{out} |\Psi_{out}^{test}\rangle_{BC}=
{1 \over{2}}\biggl(|{\theta}{\theta}\rangle_{BC}+|{\theta}{\tau}\rangle_{BC}~~~~~~~~~~~~~~~~~~~~~~~~~~~~~\nonumber\\
+|{\tau}{\theta}\rangle_{BC}+|{\tau}{\tau}\rangle_{BC}\biggr)|{\theta}{\theta}\rangle_{BC}~~~~~~~~~~\\
~~~~~~~~~~~~~~=(|{\theta}\rangle_{BC}+|{\tau}\rangle_{BC})(|{\theta}\rangle_{BC}+|{\tau}\rangle_{BC})
|{\theta}{\theta}\rangle_{BC}\nonumber.
\end{eqnarray}

If $U_{BC}$ could be implemented by a local transformation (i.e.,
if $U_{BC}=U_B\otimes U_C$) then the entanglement $E_{in}^{test}$
must equal $E_{out}^{test}$. We now show that this is not the
case.

$E_{in}^{test}$ may be calculating by noting that the
computational basis for Bob and Claire is a Schmidt basis. For any
binary string $b\in \{0,1\}^4$, the term $|b\rangle_B|b\rangle_C$
occurs in the superposition (\ref{in}) with an amplitude which
depends only on the number, $i$, of $1$'s in the binary string
$b$. Let us call the amplitude of a term with $i$ $1$'s, $\xi_i$.
Then $(\xi_0,\xi_1,\xi_2,\xi_3,\xi_4)=({1 \over 2}, {1 \over
4},0,-{1 \over 4}, -{1 \over 2})$, thus the entanglement
$E_{in}^{test}=-\sum_{i=0}^4 {4\choose i} \xi_i^2
\log_2(\xi_i^2)=3~ebits$. The final state
$|\Psi_{out}^{test}\rangle_{BC}$ clearly has
$E_{out}^{test}=2~ebits$.  We note, for future use, the the four
terms $|{\theta}{\theta}\rangle_{BC}$,
$|{\theta}{\tau}\rangle_{BC}$, $|{\tau}{\theta}\rangle_{BC}$, and
$|{\tau}{\tau}\rangle_{BC}$ that emerge from the compression add
up in such a way that their superposition is not entangled being a
product of two copies of
$(|{\theta}\rangle_{BC}+|{\tau}\rangle_{BC})$, each of them being
unentangled. The only entanglement comes from the ``factored out"
states $|{\theta}{\theta}\rangle_{BC}$. Thus we conclude that {\it
no} unitary $U_{BC}$ which acts as (\ref{basic}) can be of the
form $U_B\otimes U_C$.

Since $|\Psi_{out}^{test}\rangle_{BC}$,
$|\Psi_{in}^{test}\rangle_{BC}$ possess different amounts of
entanglement, we cannot claim that in general reversible
entanglement concentration is possible. It might be the case,
however, that in the asymptotic limit the ratio
$|E_{in}-E_{out}|/E_{in}$ goes to zero. Thus, our next step is to
find out how $|E_{in}-E_{out}|$ grows with $n$.

\section{Calculation of the entanglement difference for general
$n$.}\label{general}
 We have used a combination of analytical and
numerical techniques to find the $|E_{in}-E_{out}|$ vs. $n$
dependance.

First we derive the formula for the entanglement possessed by
$|\Psi_{in}^{test}\rangle_{BC}$ as a function of $n$ for the given
ratio $p=k/n$, where $k$ is the number of ${\tau}$'s. As the
generalization of (\ref{in}), we consider the following test
state:

\begin{equation}\label{gen_in}
 |\Psi_{in}^{test}\rangle_{BC}=\frac{1}{\sqrt{{n \choose np}}}\sum_{j}^{n \choose np}
 |P_j({\theta}^{\otimes (n-k)}{\tau}^{\otimes k})\rangle_{BC}.
 \end{equation}

\noindent where the sum runs over all possible permutations $P_j$
of $k$ $\tau$'s and $n-k$ $\theta$'s in $n$ places.

As in the case of $n=4$ the computational basis for Bob and Claire
is a Schmidt basis and for any binary string $b\in \{0,1\}^n$, the
term $|b\rangle_B|b\rangle_C$ occurs in the superposition
(\ref{gen_in}) with an amplitude which depends only on the number,
$i$, of $1$'s in the binary string $b$. Let us denote, as before,
the amplitude with $i$ $1$'s, $\xi_i$. Then the entanglement is
\begin{equation}\label{E_in_ent}
E_{in}^{test}=-\sum_{i=0}^n {n\choose i} \xi_i^2 \log_2(\xi_i^2),
\end{equation}

where

\begin{equation}\label{E_in}
\xi_i=\frac{1}{\sqrt{2^n{n \choose np}}}
\sum_{x=\max(0,i-(1-p)n)}^{\min(i,np)}(-1)^x { n-i \choose np-x}{i
\choose x}.
\end{equation}

 The entanglement $E_{out}^{test}$ is straightforward to compute
 in the case that ${n \choose k}$ is an integer power of $2$. ${n \choose
 k}=2^{N_{exact}}$, say. In this case, as in Eq. (\ref{out}),
 $|\Psi_{out}^{test}\rangle_{BC}$ is (up to normalisation) a product of
 two terms; the first term is a product of $N_{exact}$ copies of
 ($|\theta\rangle+|\tau\rangle$) and the second is a product of
 $n-N_{exact}$ copies of $|\theta\rangle$. Thus, since $|\theta\rangle+|\tau\rangle$
 is unentangled, the entanglement is
 \begin{eqnarray}\label{ent_out}
 E_{out}^{test}&=&n-N_{exact}=n-\log_2 {n \choose k}\nonumber\\
 &\sim&  n-\log_2 {n \choose np}\sim n[1-H(p)].
 \end{eqnarray}

The case when ${n \choose k}$ is not an integer power of $2$ is
more involved to analyse. However a similar situation arises in
the standard bi-partite situation \cite{entconc}. One has $n$
copies of $\sqrt {1-p}|0\rangle_A|0\rangle_B+\sqrt
{p}|1\rangle_A|1\rangle_B$ and Alice projects onto a state with a
given number, $k$, of $1$'s. In this case the issue is that
Alice's projection typically results in a state with a Schmidt
number which is not an integer power of $2$. Thus one cannot
immediately interpret the state as a certain number of singlets
held between Alice and Bob. However, as shown in \cite{entconc},
by taking batches of $n$ copies of the state ($M$ batches, say),
one can always arrange things so that the total state of the $M$
batches is as close as we like to a state with a Schmidt number an
integer power of $2$. A similar argument can be made in our
situation. Details are given in the Appendix. The result is that,
just as in the case where ${n \choose k}$ is an integer power of
$2$, $E_{out}^{test}\sim n[1-H(p)]$ with high probability.

Using these expressions we have calculated $E_{in}^{test}$ and
$E_{out}^{test}$ for different values of $p$ and $n$. Fig. 2 gives
numerical results for $E_{in}^{test}$, $E_{out}^{test}$ as a
function of $n$ for $p=0.8$. A linear dependance
$|E_{in}^{test}-E_{out}^{test}| \cong 0.466n$ is obtained. Our
calculations showed a similar behaviour for other values of $p$.
The calculated slopes $|E_{in}^{test}-E_{out}^{test}|/n$ for
several values of $p$ are plotted in Fig. \ref{slopes}. We
conclude, therefore, that since the ratio
$|E_{in}^{test}-E_{out}^{test}|/E_{in}^{test}$ is constant for
given $p$, the entanglement concentration of $|R_p\rangle$-states
cannot be performed reversibly using this standard protocol even in the asymptotic limit.\\

\begin{figure}
\epsfxsize=3.6truein \centerline{\epsffile{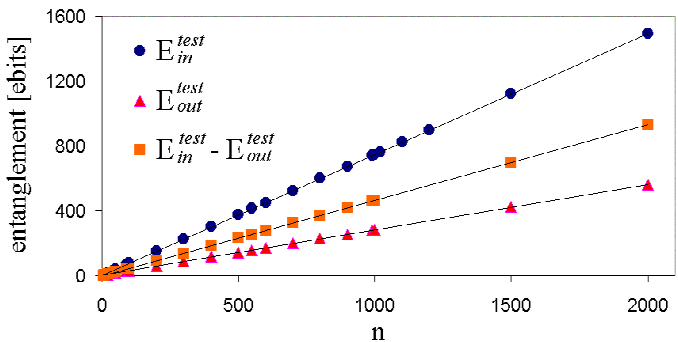}}
\vspace{0.5cm} \caption[]{Entanglement as a function of the number
of copies, $n$, for $p=0.8$. $E_{in}^{test}$ is computed
numerically from (\ref{E_in_ent}) and (\ref{E_in}); the values of
$n$ were chosen so that $np$ was an integer. For $E_{out}^{test}$
we have plotted $n-\log_2 {n \choose np}$. (The validity of this
approximation is discussed in detail in the text.)}
\label{comp_ent}
\end{figure}

\begin{figure}
\epsfxsize=3.0truein \centerline{\epsffile{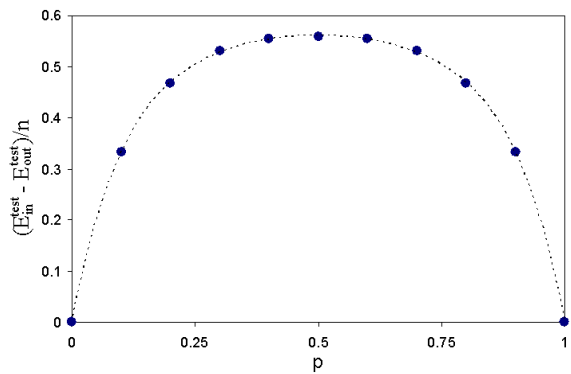}}
\caption[]{Numerical results for
$|E_{in}^{test}-E_{out}^{test}|/n$ as a function of $p$.}
\label{slopes}
\end{figure}

\section{Locking of entanglement of formation and the question of
(im)possibility of reversible concentration.}
\label{sec:local_entropies}

\cite{LPSW} provides us with necessary criteria for existence of a
reversible transformation of multi-partite entanglement: the
entropy of entanglement for all bi-partite partitions and the
relative entropy of entanglement for any two parties must remain
constant. In particular, for a three-partite state, six quantities
must be conserved: von Neumann entropies of three reduced density
matrices $S(\rho_A)$, $S(\rho_B)$ and $S(\rho_C)$, and relative
entropies of three bi-partite reduced density matrices
$E_{re}(\rho_{AB})$, $E_{re}(\rho_{BC})$ and $E_{re}(\rho_{AC})$.

 The bi-partite
entanglement $E_{A(BC)}=S(\rho_A)=nH(p)$ and
$E_{B(AC)}=E_{C(AB)}=S(\rho_B)=n$ possessed by the state
(\ref{final}) is equal to the entanglement possessed by $n$ copies
of the initial state (\ref{Tstate}). Indeed, the fact that this
hypothetical reversible concentration procedure is consistent with
these constraints, as noted in \cite{LPSW}, was a main motivation
for this paper.

We note, however, that another measure of entanglement, the
entanglement of formation $E_F$ \cite{wootters}, would not be
conserved in this hypothetical reversible transformation. Indeed,
in our case $E_F$ can be easily calculated using definitions and
results of \cite{wootters}, and for the initial state
$E_F(\rho_{BC}^{in})=nH({1\over 2}+\sqrt{p(1-p)})$, while for the
final state $E_F(\rho_{BC}^{out})=n(1-H(p))$. Here we used the
result from \cite{E_F_additivity} that $E_F$ of mixtures of the
Bell-states is additive. In general, the question of additivity of
the entanglement of formation is still open.

Although, $E_F$ is not conserved in our hypothetical
transformation, this does not automatically rule out the
transformation. Indeed the entanglement of formation
$E_F(\rho_{BC})$ can be increased by assistance from Alice.
However, an important question is how much must Alice pay (in
destroying her state) in order to increase $E_F(\rho_{BC})$. One
is tempted to assume that $|\Delta S(\rho_A)|\geq|\Delta
E_F(\rho_{BC})|$, i.e. that if Alice gains one bit of information
from her system (by measurement), then she cannot help Bob and
Claire increase their $E_F$ by more than 1 e-bit. If this were the
case it then follows that $|R_p\rangle$ cannot be reversibly
concentrated to GHZ's and EPR's {\it by any method}, because Alice
would need to destroy her entanglement with $BC$ by $n[H({1\over
2}+\sqrt{p(1-p)})+H(p)-1]$.

On the other hand, very recently, the effect of locking of
entanglement of formation \cite{locking} was discovered. For some
states $\rho_{BC}$ the entanglement of formation $E_F(\rho_{BC})$
can be increased by much more than the information received from
Alice. In principle, a single bit from Alice can result in an
increase of $E_F(\rho_{BC})$ by an arbitrarily large amount. This
offers the possibility that in our case Alice need not destroy her
entanglement with $BC$ but still allow for the required increase
in $E_F(\rho_{BC})$, and hence it might be possible to have
reversible transformation of $|R_p\rangle$-states into GHZ's and
EPR's.

We note however, that if the latter scenario were true, this would
be an example of locking of $E_F$ very different from the original
example analysed in \cite{locking}. The example in \cite{locking}
is of a specially constructed state, while in our case the state
is simply a product of GHZ's and EPR's. Furthermore, we are
considering an asymptotic situation (blocks of states) while the
example of \cite{locking} is for a single state.

\section{Discussion}
In the present paper we analyzed only one particular method, the
``the standard method", for concentrating entanglement in the case
of {\it Rohrlich} states. Using this procedure, reversible
concentration of $|R_p\rangle$ into GHZ's and singlets is not
possible. What can we conclude from this?

$\bullet$ First of all, although the state $|R_p\rangle$ was
chosen precisely because it seemed suited to concentration via the
standard protocol and it is hard to believe that other methods
could do better, that is still an open possibility. In this
context we make a number of observations:

a) We note, that in bi-partite case the task is essentially
symmetric under interchange of the roles that two parties play in
the protocol. This is not the case for three-partite interpolating
states (\ref{Tstate}). If the parties interchange their roles then
two schemes might appear as essentially different. The standard
method that we consider here is of a ``$A\rightarrow(BC)$" type,
i.e. Alice performs a collective local measurement on her side,
then reports the result to Bob and Claire, which are required to
complete the protocol by collective local unitaries on their
sides. Other methods, e.g. ``$B\rightarrow(AC)$" type, are beyond
the scope of this article.

Is the ``$A\rightarrow(BC)$" method presented here optimal amongst
all possible $A\rightarrow(BC)$ schemes? Optimality of the
standard method in bi-partite case follows from its reversibility.
Since the same method becomes irreversible when applied to
three-partite interpolating state, we cannot use the same argument
to show its optimality. Can we claim that the standard method we
use here is the most general method one can use? (If it is, then
its optimality will follow.) In the most general terms, the task
is to transform $|R_p\rangle^{\otimes n}$ into (\ref{final}). In
the asymptotic limit $|R_p\rangle^{\otimes n}$ almost entirely
lies in its typical subspace, i.e.

\begin{widetext}
\begin{equation}\label{typical1}
 |R_p\rangle^{\otimes n}\approx |\Omega\rangle=\sum_{k=np-O(\sqrt{n})}^{np+
 O(\sqrt{n})}\frac{1}{\sqrt{2^{nH}}}\sum_i|P_i(0^{\otimes (n-k)}1^{\otimes k})\rangle_A
 |P_i({\theta}^{\otimes (n-k)}{\tau}^{\otimes k})\rangle_{BC}.
 \end{equation}
\end{widetext}
Alice's local collective measurement projects $|\Omega\rangle$
into a subspace of the typical subspace, i.e. into the state
(\ref{3comp1}) with a particular value of $k$. The only way to
transform (\ref{3comp1}) to (\ref{final}) is to ``rename" the
states. This is exactly what $U_{BC}$  does. Thus, $U_{BC}$ is
most general operation needed to convert (\ref{3comp1}) into
(\ref{final}). However we  have not ruled out the possibility that
a different measurement done by Alice could project
$|\Omega\rangle$ into a state which might be converted into
(\ref{final}) using a ``cheaper" $\tilde{U}_{BC}$.

b) It is worth noting, that the standard method presented here has
features which seem undesirable in certain regimes. For example,
let us consider the situation when initially Alice, Bob and Claire
share $n$ pairs (\ref{Tstate}) which are already maximally
entangled, i.e. $p=0.5$. Clearly in this case they should not do
anything. However if they apply the concentration protocol, then
they will consume approximately $0.56n$ ebits as can be seen from
Fig. \ref{slopes}. It is not clear, however, whether this
inefficiency is an essential feature of any $|R_p\rangle$-state
concentration protocol, or only of our method.

c) Using the standard method we required that the final state
should be exactly given by singlets and GHZ's. This task demands
that Bob and Claire must use a significant amount of entanglement
to implement the required $U_{BC}$ transformation. It is possible,
however, that if we accept a final state that is only
approximately equal to a combination of singlets and GHZ's (where
the precise details of the quality of the approximation needs to
be defined appropriately), the non-locality needed by Bob and
Claire becomes negligibly small.

$\bullet$ Second, it might of course be possible that the states
$|R_p\rangle$ can be reversibly concentrated to members of
three-party MREGS other than GHZ's and singlets. This is possible
despite the fact that the concentration into EPR's and GHZ's is so
natural, both because this is what the standard method suggests as
well as the entropy considerations in \cite{LPSW}.

$\bullet$ Third, of course, it is also possible that that
reversible concentration of $|R_p\rangle$ and multi-partite states
in general is not possible.

\section{Conclusion}
We have analyzed the most natural way to concentrate multiparticle
entanglement in arguably the simplest non-trivial case. We showed
that the standard method does not work. This does not however
settle the question. There might be other methods that work, there
might be other MREGS than the one we have considered, or, of
course, concentration might fail altogether. Despite the partial
nature of our results, we feel that our analysis leads to a much
better understanding of the structure of this problem and has
implications for other areas of quantum information.

\begin{acknowledgments}
We would like to thank Andreas Winter and Tony Short for useful
comments and support. We are grateful to John Smolin, Frank
Verstraete, and Andreas Winter for sharing their recent results
\cite{svw} before publication. The authors are grateful for
support from the EU under European Commission project RESQ
(contract IST-2001-37559).
\end{acknowledgments}

\appendix
\section{The entanglement of the test state}\label{batches}
In the text in Sec. \ref{general} we noted that when Alice
measures her system, she will find $k$ $1$'s out of a total of
$n$, but that ${n \choose k}$ may not be an integer power of $2$.
The computation of $E_{out}^{test}$ is simple when ${n \choose k}$
is a power of $2$; and with high probability $k$ will be close to
$np$ and therefore $E_{out}^{test}\sim n[1-H(p)]$. If ${n \choose
k}$ is not an integer power of $2$ then we need to adapt the
procedure in order to get direct product of perfect GHZ's.
 Here we show that when we do so, the leading order
behaviour is still that $E_{out}^{test}\sim n[1-H(p)]$ with high
probability.

We will follow the standard bi-partite concentration procedure
\cite{entconc} and take $M$ batches with $n$ states in each batch.
Measurement of the $i$-th batch yields $k_i$ one's. Let $D_M$
denote the accumulated product $\Pi_{i=1}^M {n \choose k_i}$. We
continue measuring batches of $n$ states until $D_M$ is in the
interval $[2^l, 2^l(1+\epsilon)]$, for some integer $l$ and some
small fixed $\epsilon$. The expected number of batches is
$1/\epsilon$, and the expected total number of states in the
ensemble is, therefore, $N=n/\epsilon$.

$U_{BC}$ acts on the complete set of batches. Each term is a
string of $\log_2 D_M$ qubit pairs; each qubit pair is in the
state $|\theta\rangle$ or the state $|\tau\rangle$. $U_{BC}$
transforms (``compresses") each string to one in which the
trailing qubits are all in the state $|\theta\rangle$ [cf Eqn.
(\ref{basic})].

As in the body of the text, we will bound the entanglement in
$U_{BC}$ by considering its action on a test state. We take as a
test state the tensor product of the bi-partite test states used
in the text, one for each batch of $n$ pairs:
\begin{equation}
|\Theta_{in}^{test}\rangle=|\Psi_{in}^{test}(k_1)\rangle\otimes|\Psi_{in}^{test}(k_2)\rangle\otimes
...\otimes|\Psi_{in}^{test}(k_M)\rangle.
\end{equation}
The entanglement of $|\Theta_{in}^{test}\rangle$ is the sum of the
entanglement of all states $|\Psi_{in}^{test}(k_i)\rangle_{BC}$,
which in the asymptotic limit is just $n$ times the entanglement
of a single $|\Psi_{in}^{test}(k)\rangle_{BC}$ with $k=np$, which
we calculated numerically in the body of the text.

$|\Theta_{out}^{test}\rangle$ is a product of two terms
\begin{equation}\label{theta_out}
|\Theta_{out}^{test}\rangle=|\Gamma_M\rangle
\otimes|{\theta}{\theta}{\theta}...{\theta}\rangle,
\end{equation}
where
\begin{eqnarray}\label{Gamma_M}
|\Gamma_M\rangle=&{1 \over \sqrt{\gamma}} \biggl[
|{\theta}\rangle\biggl(|{\theta}.....{\theta}{\theta}\rangle+|{\theta}.....{\theta}{\tau}\rangle
+...+|{\tau}.....{\tau}{\tau}\rangle\biggr)\nonumber\\&
+|{\tau}\rangle\biggl(|{\theta}........{\theta}\rangle+...+
|{\theta}{\theta}...{\theta}{\tau}...\rangle\biggr)\biggr],~~~~
\end{eqnarray}
where $\gamma$ is a normalisation factor and equals the total
number of terms and lies between $2^l$ and $2^l(1+\epsilon)$, i.e.
$\gamma=2^l(1+\epsilon^{'})$, where $0\leq \epsilon^{'}\leq
\epsilon$. The number of $\theta$'s in the second term of
(\ref{theta_out}) will be close to $Mn(1-H)=N(1-H)$ with high
probability.

Thus $|\Gamma_M\rangle$ can be written
\begin{equation}
|\Gamma_M\rangle=\sqrt{1 \over
{1+\epsilon^{'}}}|\phi_1\rangle+\sqrt{\epsilon^{'} \over
{1+\epsilon^{'}}} |\phi_2\rangle,
\end{equation}
where
\begin{eqnarray}
|\phi_1\rangle={1 \over
\sqrt{2^l}}|{\theta}\rangle\biggl(|{\theta}.....{\theta}{\theta}\rangle+...+|{\tau}.....{\tau}{\tau}\rangle\biggr)
\end{eqnarray}
contains the first $2^l$ terms and
\begin{eqnarray}
|\phi_2\rangle={1 \over
\sqrt{\epsilon^{'}2^l}}|{\tau}\rangle\biggl(|{\theta}........{\theta}\rangle+...+
|{\theta}{\theta}...{\theta}{\tau}...\rangle\biggr)
\end{eqnarray}
the remaining $\epsilon^{'} 2^l$ terms ($|\phi_1\rangle$ and
$|\phi_2\rangle$ are orthogonal and normalised).

We now use the fact \cite{bound} that for any two bi-partite pure
orthogonal states $|\phi_1\rangle$ and $|\phi_2\rangle$,
  the entanglement  of the
  superposition $\alpha|\phi_1\rangle+\beta
|\phi_2\rangle$ satisfies:
\begin{equation}\label{ent_bound}
 E(\alpha|\phi_1\rangle+\beta
|\phi_2\rangle)\leq 2[|\alpha|^2 E(\phi_1)+|\beta|^2
E(\phi_2)+H(|\alpha|^2)].
\end{equation}

The entanglement of $|\phi_1\rangle$ is 1 ebit, while the
entanglement of $|\phi_2\rangle$ is at most $N$.

Also ${\epsilon^{'} \over {1+\epsilon^{'}}} \leq\epsilon$, ${1
\over {1+\epsilon^{'}}}\leq 1$, and $H(\epsilon)\leq 1$, thus
(\ref{ent_bound}) shows that the entanglement of
$|\Gamma_M\rangle$ satisfies
\begin{equation} E(|\Gamma_M\rangle)\leq
2[1+\epsilon N+ 1]=2(\epsilon N+2).
\end{equation}

Thus the entanglement per batch that $|\Gamma_M\rangle$
contributes is $ 2(\epsilon N+2)\epsilon \sim 2n\epsilon$ (recall
that the expected number of batches is $1/\epsilon$).  However, as
we have observed earlier, the expected number of $\theta$'s in the
second term in (\ref{theta_out}) is $N(1-H)$, so the entanglement
per batch associated with this second term is expected to be
$N(1-H)\epsilon \sim  n(1-H)$.  Thus the entanglement of
$|\Gamma_M\rangle$ is negligible, just as it was when ${n \choose
k}$  was an integer power of 2, and so the expected entanglement
per batch will be $n(1-H)$.


\end{document}